\documentclass[12pt, a4paper]{article}
\usepackage[english]{babel}
\usepackage{graphicx}
\begin{document}
\title{\bf\boldmath THE PROCESS 
$e^+e^-\to\omega\pi^0\to\pi^0\pi^0\gamma$ UP TO 1.4 GEV.}
\author{
M.N.Achasov, K.I.Beloborodov, A.V.Berdyugin,\\
A.G.Bogdanchikov, A.V.Bozhenok, D.A.Bukin,S.V.Burdin,\\
V.B.Golubev, T.V.Dimova, A.A.Drozdetsky, 
V.P.Druzhinin\thanks{e-mail: druzhinin@inp.nsk.su},\\
M.S.Dubrovin, V.N.Ivanchenko, A.A.Korol, S.V.Koshuba,\\
G.A.Kukartsev, I.N.Nesterenko, E.V.Pakhtusova, E.A.Perevedentsev,\\
V.M.Popov, A.A.Salnikov, S.I.Serednyakov, V.V.Shary, Yu.M.Shatunov,\\
V.A.Sidorov, Z.K.Silagadze, A.N.Skrinsky, A.G.Skripkin,\\
Yu.V.Usov, A.A.Valishev, A.V.Vasiljev, Yu.S.Velikzhanin
\vspace*{3mm}\\ Budker Institute of Nuclear Physics,  630 090,
Novosibirsk, Russia}
\date{}
\maketitle
\begin{abstract}
The cross section of the $e^+e^-\to\omega\pi^0\to\pi^0\pi^0\gamma$ reaction
was measured by the SND detector at VEPP-2M $e^+e^-$ collider in the
energy range from
threshold up to 1.4 GeV. Results of the cross section 
fitting by the sum of $\rho$, $\rho^{\prime}$ and $\rho^{\prime\prime}$
contributions are presented.
\end{abstract}
\section{Introduction}
The process of $e^+e^-$ annihilation into hadrons in the 1--2 GeV 
energy region
is an important source of information about excited
states of light vector mesons $\rho$, $\omega$ and $\phi$.
Parameters of these states are still not well established
mainly due to poor accuracy of existing experimental data.
Recently new data 
in this energy region became available. The cross sections of the reactions 
$e^+e^-\to 3\pi$
\cite{3pi} and $e^+e^-\to 4\pi$ \cite{cmd-4pi} were measured
at VEPP-2M collider. Accuracy was also significantly improved
in the recent measurements of $\tau\to 2\pi\nu_\tau$ 
and $\tau\to 4\pi\nu_\tau$
spectral functions \cite{cleo-2pi,cleo-4pi,aleph} related to corresponding
$e^+e^-$ annihilation cross sections by the CVC hypothesis \cite{Tsai}.

The reaction $e^+e^-\to\omega\pi^0$ is one of the dominant processes in
the energy range from 1 to 2 GeV. The PDG values for the mass and width of the
$\rho^\prime(1450)$ meson
\cite{pdg} are based mainly on phenomenological studies of this process  
\cite{Clegg}. The most precise measurements of the 
$e^+e^-\to\omega\pi^0$ cross section  
and $\tau\to \omega\pi\nu_\tau$ 
spectral function were done in \cite{cmd-4pi,dm2-4pi,cleo-4pi}. 
All these experiments addressed the $4\pi$ final state into which other 
intermediate states, for example $a_1\pi$, contribute significantly.
In this work the $e^+e^-\to\omega\pi^0$ reaction was studied
in the $\pi^0\pi^0\gamma$ final state where other contributions are
much smaller.
This allows to avoid systematic errors
inherent to the $e^+e^-\to 4\pi$ channel due to
non-trivial background
subtraction, which must take into account interference effects.
The process
$e^+e^-\to\omega\pi^0\to\pi^0\pi^0\gamma$ was studied for the first time 
in \cite{nd} where 20\% statistical accuracy was achieved.

\section{Detector and experiment}
SND is a general purpose non-magnetic detector \cite{SND}. Its main
part is a three-layer scintillation electromagnetic calorimeter consisted
of 1630 NaI(Tl) crystals with solid angle coverage about 90\% of 4$\pi$.
The energy resolution of the calorimeter for photons is
$\sigma_E/E=4.2\%/\sqrt[4]{E(\mbox{GeV})}$, the angular
resolution is about
1.5$^{\circ}$. The directions of charged particles are measured by
two coaxial cylindrical drift chambers covering 95\% of $4\pi$ solid angle.              

The analysis presented in this work is based on data recorded in 
1997--1999 in two separate energy regions: 920--980 MeV and
1040--1380 MeV. Analysis of the $\phi$-resonance region (980--1040 MeV) was
published earlier
\cite{phi-omp}. In the first energy region the total
integrated luminosity of 1.5 $\mbox{pb}^{-1}$
was collected at 6 energy points. The region above
the $\phi$ meson was scanned with a 10 MeV step. Total integrated luminosity 
accumulated in this
region is about 9 $\mbox{pb}^{-1}$. The luminosity was measured with 
a systematic uncertainty of 3\%
using $e^+e^-\to e^+e^-$ and $e^+e^-\to\gamma\gamma$ reactions. 
\section{Event selection} 
For primary selection of     
\begin{equation}
 e^{+}e^{-} \to\omega\pi^0\to\pi^0\pi^0\gamma\label{ompn} 
\end{equation}
events the following criteria were applied:
\begin{itemize}
\item five or more photons and no charged tracks are found in an event;
\item the energy deposition in the calorimeter is more than $0.7 E$;
\item the total momentum of an event measured by the calorimeter is less than 
$0.15 E$;
\end{itemize}
where $E$ is a center of mass energy of $e^+e^-$ pair.
The main sources of background surviving these cuts are           
QED processes
\begin{equation}
 e^{+}e^{-} \to 2\gamma,\,3\gamma.\label{3gam}
\end{equation}
with extra photons either from the the beam background or
splitting of electromagnetic showers in the calorimeter.

For each event satisfying primary selection criteria the kinematic
fitting assuming
$e^{+}e^{-} \to \pi^0\pi^0\gamma\to 5\gamma$ hypothesis was performed.
As a result, two parameters: $\chi_{\pi\pi\gamma}$ --- the
$\chi^2$ of the fit and $M_{\pi\gamma}$ --- the $\pi^0$ recoil mass closest to
that of $\omega$ meson were evaluated.

\begin{figure}[t]
\begin{minipage}[t]{0.47\textwidth}
\includegraphics[width=0.98\textwidth]{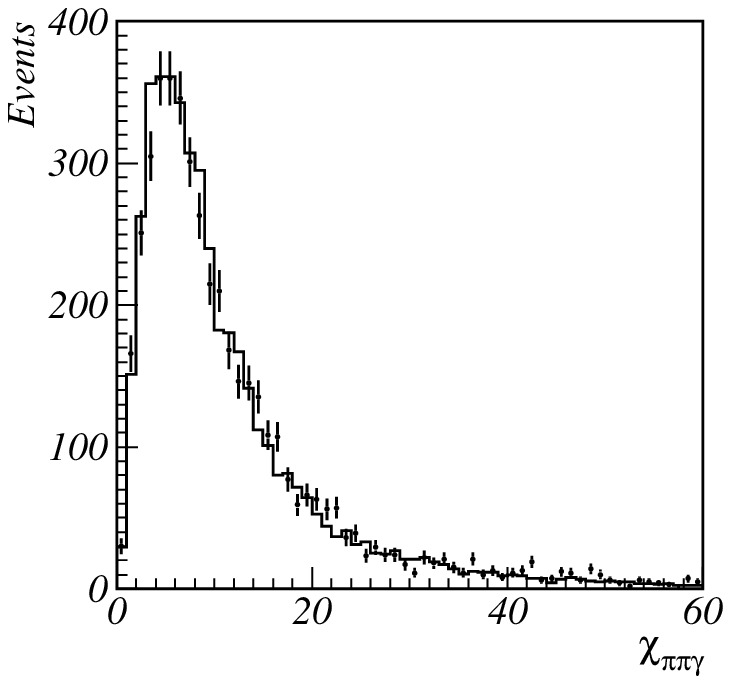}
\caption{\small The $\chi_{\pi\pi\gamma}$ distribution
for the experimental (points with error bars) and simulated (the histogram)
events of the process (\ref{ompn}). }
\label{f1}
\end{minipage}
\hfill
\begin{minipage}[t]{0.47\textwidth}
\includegraphics[width=0.98\textwidth]{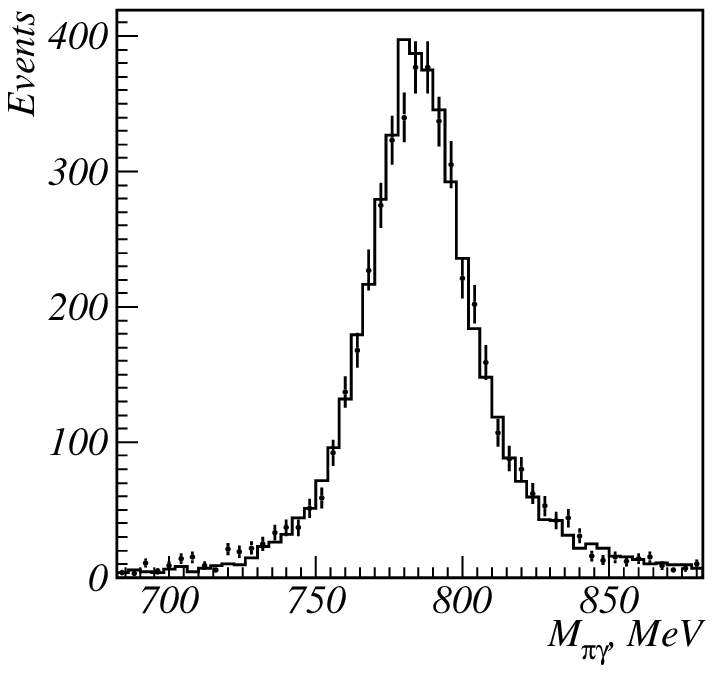}
\caption{\small The $M_{\pi\gamma}$ distribution
for the experimental (points with error bars) and simulated (the histogram)
events of the process (\ref{ompn}). }
\label{f2}
\end{minipage}
\end{figure}
The $\chi_{\pi\pi\gamma}$ distribution for the events 
with $|M_{\pi\gamma}-782| < 50$ and
the $\pi^0$ recoil mass spectrum for the events with $\chi_{\pi\pi\gamma}<30$
are plotted in figs.\ref{f1} and \ref{f2}.
Good agreement between experimental distributions
and simulation of the process (\ref{ompn}) shows that
there are no other
significant contributions in the selected event sample. 
For final event selection
the following cuts were applied:
\begin{equation}
\chi_{\pi\pi\gamma} < 30,\,|M_{\pi\gamma}-782| < 50,\label{seln}
\end{equation}
which reject the QED background (\ref{3gam}) almost completely.
In order to estimate residual background the experimental
$\pi^0$ recoil
mass spectrum was fitted by the sum of simulated spectrum of 
the process (\ref{ompn})
and a linear background. As a result the estimated total number of
background events did not exceed  1.5\% of all 
selected events. This value was taken as a
systematic error of the measured cross section of the process 
(\ref{ompn}) related to the residual background.

Experimental data were collected in 40 energy points. In the energy
region under study the cross section changes slowly so it was possible to
reduce the number of energy points combining the neighboring ones.
Resulting energies, their standard deviations, integrated luminosities ($IL$)
and numbers of
selected events ($N$) are listed in the Table \ref{tab1}. The detection 
efficiency for the process (\ref{ompn}) was determined by simulation based on
a formula from \cite{ompn} which takes into account finite width of
$\omega$-meson.
In the energy range from 
1050 to 1400 MeV the efficiency was found to be
constant and equal to 40\%.
Near  the $e^+e^-\to\omega\pi^0$ threshold
the number of events with the $\pi^0$ recoil
mass below $\omega$-meson mass increases sharply and the efficiency goes
down.
The energy dependence of the detection efficiency is presented in the 
Table~\ref{tab1}.                
\section{Fitting of the cross section}
\begin{table}[t]
\small
\caption{\small Energy, its standard deviation, integrated luminosity,
detection efficiency, number of events, radiative correction and
cross section of the process 
$e^+e^-\to\omega\pi^0\to\pi^0\pi^0\gamma$}
\begin{center}
\begin{tabular}{|c|c|c|c|c|c|c|c|}
\hline
$E$, MeV& $\Delta E$, MeV & $IL$, nb & $\varepsilon$ & $N$ & $\delta$ &
$\sigma_0$, nb\\
\hline
 920 & 0.2 & 328 & 0.169 &   3 & -0.148 & 0.06$\pm$0.04\\
 940 & 0.2 & 289 & 0.312 &  15 & -0.168 & 0.20$\pm$0.05\\
 954 & 2.0 & 496 & 0.332 &  39 & -0.157 & 0.28$\pm$0.05\\
 973 & 4.3 & 373 & 0.352 &  68 & -0.140 & 0.60$\pm$0.07\\
1020 & --  & --  &  --   &  -- &   --   & 0.74$\pm$0.02\\
1045 & 5.0 & 152 & 0.390 &  51 & -0.100 & 0.96$\pm$0.13\\
1063 & 4.4 & 371 & 0.390 & 124 & -0.093 & 0.95$\pm$0.09\\
1081 & 3.5 & 666 & 0.390 & 253 & -0.086 & 1.07$\pm$0.07\\
1102 & 3.8 & 524 & 0.390 & 180 & -0.080 & 0.96$\pm$0.07\\
1123 & 4.4 & 420 & 0.390 & 171 & -0.074 & 1.13$\pm$0.09\\
1142 & 4.0 & 358 & 0.390 & 142 & -0.070 & 1.09$\pm$0.09\\
1160 & 0.4 & 316 & 0.390 & 158 & -0.066 & 1.37$\pm$0.11\\
1183 & 4.6 & 587 & 0.390 & 280 & -0.061 & 1.30$\pm$0.08\\
1203 & 4.4 & 569 & 0.390 & 284 & -0.057 & 1.36$\pm$0.08\\
1223 & 4.6 & 465 & 0.390 & 232 & -0.054 & 1.35$\pm$0.09\\
1243 & 4.8 & 562 & 0.390 & 309 & -0.050 & 1.48$\pm$0.08\\
1266 & 5.1 & 397 & 0.390 & 241 & -0.046 & 1.63$\pm$0.11\\
1285 & 5.2 & 492 & 0.390 & 268 & -0.043 & 1.46$\pm$0.09\\
1304 & 5.0 & 459 & 0.390 & 249 & -0.040 & 1.45$\pm$0.09\\
1325 & 5.0 & 516 & 0.390 & 301 & -0.038 & 1.55$\pm$0.04\\
1344 & 5.8 & 676 & 0.390 & 387 & -0.036 & 1.52$\pm$0.08\\
1363 & 4.6 & 857 & 0.390 & 497 & -0.034 & 1.54$\pm$0.07\\
1380 & 0.5 & 470 & 0.390 & 234 & -0.033 & 1.32$\pm$0.09\\
\hline
\end{tabular}
\end{center}
\label{tab1}
\end{table}
The visible cross section $\sigma_{vis}=N/IL$ is related with the Born
cross section of $e^+e^-\to\omega\pi^0\to\pi^0\pi^0\gamma$ process as
\begin{equation}
\sigma_{vis}(E) = \varepsilon(E)\sigma_0(E)(1+\delta(E)),\label{csvis}
\end{equation}
where $\delta(E)$ is a radiative correction calculated 
according to \cite{radc}.
Radiative corrections for the different energies and obtained cross
section values are listed in the Table \ref{tab1}.
The cross section at $E=m_\phi$ was taken from \cite{phi-omp}.
Only statistical errors are shown in the table. The systematic error
includes the error of the
luminosity measurement (3\%),
the detection efficiency error (4\%), 
possible background contribution (1.5\%),
and the error of radiative correction (1\%). 
The total systematic error was estimated to be 5\%.

\begin{figure}[t]
\includegraphics[width=0.95\textwidth]{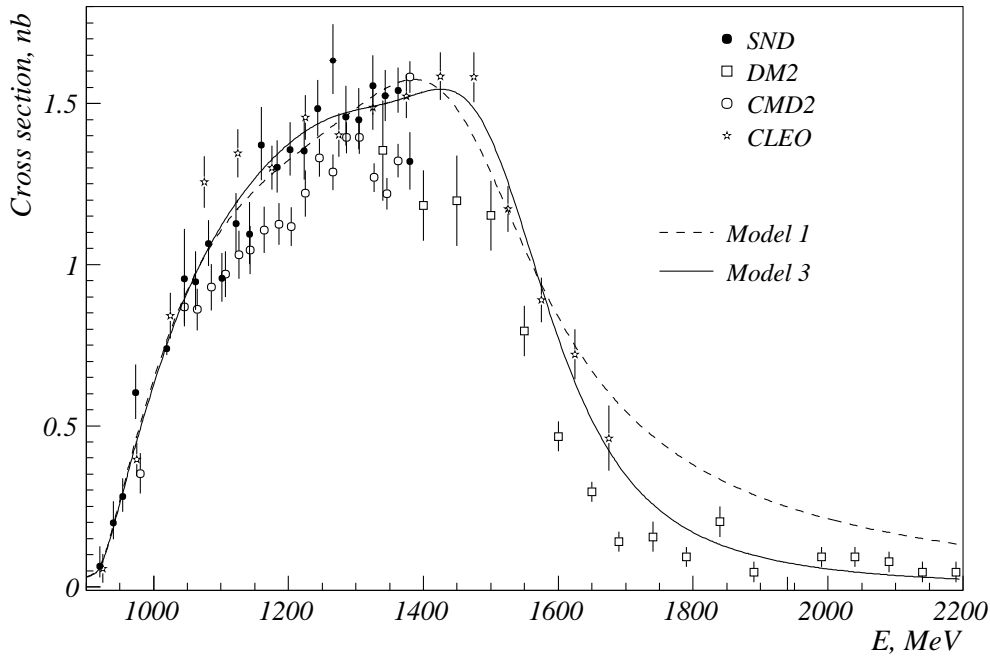}
\caption{\small The cross section of the reaction
$e^+e^-\to\omega\pi^0\to\pi^0\pi^0\gamma$. The results
of the SND (this work), DM2 \cite{dm2-4pi}, CMD \cite{cmd-4pi},
CLEO \cite{cleo-4pi} experiments are shown. Curves are
results of fitting to the data in model 1 and 3 with $R=0$.}
\label{f3}
\end{figure}

Our results in comparison with the most precise
CMD-2~\cite{cmd-4pi}, CLEO~\cite{cleo-4pi}, and DM2~\cite{dm2-4pi}
measurements are shown in Fig.\ref{f3}.
The cross sections from \cite{cmd-4pi,dm2-4pi} measured
in the $\pi^+\pi^-\pi^0\pi^0$ channel were recalculated using the PDG 
value
of $B(\omega\to\pi^0\gamma)$\cite{pdg}. The cross section from \cite{cleo-4pi}
was obtained from the $\tau\to\omega\pi\nu_\tau$ spectral function
assuming the CVC hypothesis \cite{Tsai}. The CLEO results are in good 
agreement with ours while the CMD-2 measurements are about 10\%
lower, although
the difference observed is smaller than the 15\%
systematic error quoted in
\cite{cmd-4pi}. There is a significant difference between the results of 
DM2~\cite{dm2-4pi} and CLEO~\cite{cleo-4pi}. For the cross 
section fitting we used our data together with the data from CLEO.

The energy dependence of the process (\ref{ompn}) cross section was written
as a sum of contributions from $\rho(770)$ and its excitations
$\rho^\prime$ and $\rho^{\prime\prime}$:
\begin{equation}
\sigma_0(E)=\frac{4\pi\alpha^2}{E^3}\biggl(\frac{g_{\rho\omega\pi}}
{f_\rho}\biggr)^2\biggl |\frac{m_\rho^2}{D_\rho}C_{\rho\omega\pi}+
A_1\frac{m_{\rho^\prime}^2}{D_{\rho^\prime}}C_{\rho^\prime\omega\pi}+
A_2\frac{m_{\rho^{\prime\prime}}^2}{D_{\rho^{\prime\prime}}}
C_{\rho^{\prime\prime}\omega\pi}\biggr |^2 P_f(E).
\label{ompcs}                         
\end{equation}
Here $\alpha$ is a fine structure constant and
$g_{\rho\omega\pi}$ is a $\rho\to\omega\pi$ coupling
constant. The $f_\rho$ coupling constant  was calculated
from the $\rho\to e^+e^-$ decay width:
$\Gamma_{\rho ee}=4\pi m_\rho \alpha^2/3f_\rho^2$. The expression
${m_\rho^2}/{D_\rho}$ represents $\rho$-meson Breit-Wigner amplitude with
$D_\rho=m_\rho^2-E^2-iE\Gamma_\rho(E)$, where $m_\rho$ and $\Gamma_\rho(E)$
denote the $\rho$ meson mass and energy-dependent total width respectively.
The real parameters
$A_i=g_{\rho_i\omega\pi}/g_{\rho\omega\pi}\cdot f_\rho/f_{\rho_i}$
are the ratios of the coupling constants of different $\rho$ states.
The factor $P_f(E)$ describes the energy dependence of the final state 
phase space. In the case of infinitely narrow $\omega$ resonance
$P_f(E)=1/3\cdot q_\omega^3B(\omega\to\pi^0\gamma)$, where
$q_\omega$ is an $\omega$-meson momentum. This approximation is good
for the energy range above 1050 MeV, but at energies close to
the $e^+e^-\to\omega\pi^0$ threshold
it is more adequate to use precise formula
taking into account finite width of the $\omega$ meson \cite{ompn}.
The Blatt-Weisskopf factors $C_{\rho_i\omega\pi}$ restricting fast 
growth of the partial widths, were taken in the form \cite{Clegg}:
\begin{equation}
C_{\rho\omega\pi}=\frac{1}{1+(Rq_\omega(E))^2},\:
C_{\rho_i\omega\pi}=
\frac{1+(Rq_\omega(m_{\rho_i}))^2}
     {1+(Rq_\omega(E))^2},\; i=\rho^\prime,\rho^{\prime\prime},
\label{Compi}
\end{equation}
The range parameter $R$ was assumed to be the same for $\rho$,
$\rho^\prime$ and $\rho^{\prime\prime}$ mesons.
The energy dependence of the $\rho(770)$ total width 
was expressed as:
\begin{equation}
\Gamma_{\rho}(E)=\Gamma_{\rho}(m_\rho)
\biggl(\frac{m_\rho}{E}\biggr)^2
\biggl(\frac{q_\pi(E)}{q_\pi(m_\rho)}\biggr)^3 C_{\rho\pi\pi}^2+
\frac{g_{\rho\omega\pi}^2}{12\pi}\, q_\omega^3(E)\, C_{\rho\omega\pi}^2,
\label{grho}
\end{equation}
where $q_\pi$ is the pion momentum, $C_{\rho\pi\pi}$ is the Blatt-Weisskopf
factor for $\rho\to\pi^+\pi^-$ decay \cite{Pisut}:
\begin{equation}
C_{\rho\pi\pi}(E)=\sqrt{\frac{1+(Rq_\pi(m_{\rho}))^2}
                   {1+(Rq_\pi(E))^2}}.
\label{Cpipi}
\end{equation}
There is no generally accepted  description of the shapes of
the broad excited states $\rho^\prime$ and $\rho^{\prime\prime}$.
In  \cite{Clegg} constant total width were
assumed, while in \cite{Achasov-rho} the total widths varied with energy
as a sum of the partial widths of all main decay modes.
\begin{table}[t]
\small
\caption{\small The fit results for various models described in the text.}
\begin{center}
\begin{tabular}{|c|c|c|c|c|}
\hline
& Model 1& Model 2& Model 3 & Error\\
\cline{2-5}
$R$, GeV$^{-1}$&0--2&0--2&0--2&\\
$g_{\rho\omega\pi}$, GeV$^{-1}$&16.1--13.3&15.6--13.2&15.6--12.9&0.3--0.4\\
$m_{\rho^\prime}$, MeV&1460--1520& -- &$\equiv1400.$&10\\
$\Gamma_{\rho^\prime}$, MeV&380--500& --&$\equiv500.$&20--30\\
$A_1$&-(0.22--0.24)& $\equiv0.$ &-(0.04--0.08)&0.01--0.02 \\
$m_{\rho^{\prime\prime}}$, MeV&--&1710--1580&1620--1550&15--20\\
$\Gamma_{\rho^{\prime\prime}}$, MeV&--&1040--490&580--350&70--20\\
$A_2$&$\equiv0.$&-(0.22--0.20)&-(0.18--0.13)&0.1\\
$\chi^2/ND$&(52--48)/35&(47--48)/35&(43--44)/34&\\                                   
\hline
\end{tabular}
\end{center}
\label{tab2}
\end{table}
We considered the both approaches in order to understand model
dependence of the fit parameters.

The fit results obtained for three classes of models 
are listed in the Table~\ref{tab2}. The fit parameters are
sensitive to the variation of the range parameters R in the
Blatt-Weisskopf factors. In the Table~\ref{tab2} we show the intervals of 
the fit parameter
variation when $R$ ranges from 0 to 2 GeV$^{-1}$.
The typical errors of the parameters obtained for each specific
model are listed in the fourth column of the Table~\ref{tab2}.

In the model 1 the constant widths of $\rho^\prime$ and
$\rho^{\prime\prime}$ were assumed. 
The mass and width of the $\rho^{\prime\prime}$ meson were fixed
to their PDG values \cite{pdg}:
$m_{\rho^{\prime\prime}}=1700 \mbox{ MeV}$,
$\Gamma{\rho^{\prime\prime}}=235 \mbox{ MeV}$.
The obtained value of the $\rho^{\prime\prime}$ amplitude $A_2$ is
compatible with zero, so the final fit result for the model 1 is
given for $A_2$ fixed to zero. Let us discuss the choice of
an upper boundary 
for the $R$ parameter. Small values $g_{\rho\omega\pi}<13$ 
GeV$^{-1}$ found for $R>2$ GeV$^{-1}$ are in conflict with the QCD
sum rules estimation 16 GeV$^{-1}$ \cite{Lublinsky} and
experimental value of 14.4 GeV$^{-1}$, obtained from the
$\omega\to 3\pi$ decay width assuming that $\omega\to\rho\pi$ mechanism
dominates in this decay.
Only for $R\sim0$ the extracted mass and width of the $\rho^\prime$ meson
are compatible with their PDG values\cite{pdg}. However in this case
the fit yields the lowest confidence level of $P(\chi^2)=3\%$.

Models 2 and 3 take into account energy dependence
of the total widths of $\rho^\prime$ and $\rho^{\prime\prime}$ mesons.
Since branching ratios of $\rho^{\prime}$
and $\rho^{\prime\prime}$ decays are practically unknown \cite{pdg}
the energy dependence of the sum of all rapidly
growing partial widths of their multihadron decays was approximated
as the energy dependence of $\rho_i\to\omega\pi$ width:
\begin{eqnarray}
\Gamma_{\rho_i}(E)&=&\Gamma_{\rho_i}(m_{\rho_i})\biggl [
(1-B_{\rho_i\to\pi\pi})
\biggl (\frac{q_\omega(E)}{q_\omega(m_{\rho_i})}\biggr)^3
C_{\rho_i\omega\pi}^2\nonumber\\ 
&+&B_{\rho_i\to\pi\pi}
\biggl (\frac{m_{\rho_i}}{E}\biggr)^2
\biggl (\frac{q_\pi(E)}{q_\pi(m_{\rho_i})}\biggr)^3
C_{\rho_i\pi\pi}^2\biggr],
\label{grhop}
\end{eqnarray}
For the $\rho^{\prime\prime}\to 2\pi$ branching fraction we use the 
theoretical estimation: $B_{\rho^{\prime\prime}\to\pi\pi}=10\%$ \cite{Barnes}.
For the $\rho^\prime$ meson the value $B_{\rho^\prime\to\pi\pi}=50\%$ was
chosen which, we think, reflects experimental situation 
more correctly than the theoretical prediction $25\%$ \cite{Barnes}.
The only difference between the models 1 and 2 which both consider
only one excited $\rho$ state is that the model 2 assumes its
total width energy dependent. But the fit results
for these two models are quite different. Particularly the 
mass found for the model 2 is close to the PDG value of the
$\rho^{\prime\prime}$-meson mass.
On the other hand there is a definite signal of the $\rho^\prime$
meson with a mass of 1320--1400~MeV and 400--500~MeV width in the pion
form factor data~\cite{cleo-2pi,aleph}.
Therefore we also considered the model 3 with
two excited $\rho$ states, in which the mass and width of the $\rho^\prime$ 
meson were fixed to $m_{\rho^{\prime}}=1400 \mbox{ MeV}$ and
$\Gamma_{\rho^{\prime}}=500 \mbox{ MeV}$.
This model yields the best fit to the experimental data:
$P(\chi^2)\approx13\%$. Fitting curves corresponding
to the models 1 and 3 with $R=0$ are shown in figure ~\ref{f3}.

The main conclusions from the analysis of the fit results are the following.
Fitting of the same experimental data by
models with fixed and energy-dependent total widths of the excited states
yields quite different parameters of these states.
This is caused by strong energy dependence
of the phase space for the main decay modes of $\rho^\prime$ and
$\rho^{\prime\prime}$ mesons and this effect should be
taken into account in the fitting of experimental data.
Satisfactory description of the experimental cross section
was obtained in the
model with two excited states with the masses $m_{\rho^\prime}=1400$ MeV
and $m_{\rho^{\prime\prime}}\approx1600$ MeV in which contribution
of the higher state dominates.
However this result contradicts the
theoretical expectation \cite{Barnes, Isgur}, where
$\rho^\prime$ and $\rho^{\prime\prime}$ are considered
as $2S$ and $1D$ $q\bar q$ states respectively and the larger contribution of
the lower $2S$ excitation was predicted.
Thus, with the new experimental data the situation in the isovector sector
remains unclear. The main problem for data analysis is the absence 
of consistent phenomenological description of the shapes 
of broad resonances with strong energy dependence of partial widths.

\section{Summary}
In this work the cross section of the 
$e^+e^-\to\omega\pi^0\to\pi^0\pi^0\gamma$
reaction was measured from the threshold up to 1.4 GeV with a 5\%
systematic accuracy. This is the most precise measurement of the 
$e^+e^-\to\omega\pi^0$ cross section
in this energy range. Our data are in a good agreement with the CLEO 
measurement of $\tau\to\omega\pi\nu_\tau$ spectral
function \cite{cleo-4pi}. The combined fit to our and CLEO data was
performed in the vector meson dominance model taking into account
the contributions 
of $\rho$, $\rho^\prime$ and $\rho^{\prime\prime}$ states. The
experimental data can be reasonably well approximated assuming 
existence of only one excited state but its mass is 
strongly model-dependent  and varies from 1460 to 1700 MeV for
different descriptions of the resonance shape.
The best fit to the data was obtained for the model with two excited states
in which the higher state with $m_{\rho^{\prime\prime}}\approx1600$
MeV dominates.

This work is supported in part by the Russian Fund for Basic                    
Researches (grants No. 99-02-16815, 99-02-17155) and STP ``Integration''
(grant No. 274).

\end{document}